# Theoretical Economics as Successive Approximations of Statistical Moments

Victor Olkhov

Independent, Moscow, Russia

victor.olkhov@gmail.com

ORCID: 0000-0003-0944-5113

**April 15, 2024**

### Abstract

This paper studies the links between the descriptions of macroeconomic variables and statistical moments of market trade, price, and return. The randomness of market trade values and volumes during the averaging interval $\Delta$ results in the random properties of price and return. We describe how averages and volatilities of price and return depend on the averages, volatilities, and correlations of market trade values and volumes. The averages, volatilities, and correlations of market trade, price, and return can behave randomly during the long interval $\Delta_2 >> \Delta$. To describe their statistical properties during the long interval $\Delta_2$, we introduce the secondary averaging procedure of trade, price, and return. We explain why, in the coming years, predictions of market-based probabilities of price and return will be limited by Gaussian distributions. We discuss the roots of the internal weakness of the commonly used hedging tool, Value-at-Risk, that cannot be solved and remains the source of additional risks and losses. One should consider theoretical economics as a set of successive approximations, each of which describes the next array of the n-th statistical moments of market trades, price, return, and macroeconomic variables, which are repeatedly averaged during the sequence of increasing time intervals.

---

 This research received no support, specific grants, or financial assistance from funding agencies in the public, commercial, or nonprofit sectors. We welcome valuable offers of grants, support, and positions.

## 1. Introduction

Studies of economic theory, starting at least with the *Essay* by Cantillon (1730), are presented in numerous papers (Hicks, 1937; Schumpeter, 1939; Neumann, 1945; Solow, 1956; Leontief, 1973; Sargent, 1979; Blaug, 1985; Greenwald and Stiglitz, 1987; Romer, 1996; Krueger, 2002; Kurz and Salvadori, 2003; Wickens, 2008; Vines and Wills, 2018). Any review of the current state of economic theory should discuss the results of hundreds of researchers.

To avoid that impossible task, we direct our efforts to the investigation of the general framework of economic theory. We consider the links between the composition of macroeconomic variables, variables of economic agents, and market transactions. We believe that market trade as the only origin of economic development and the main root of economic uncertainty and stochasticity. The randomness of market trades can be explained by many factors, such as the uncertainty of agents' expectations, economic or political shocks, etc. A lot of hidden reasons cause the stochasticity of market trade. However, all these hidden factors result in the irregularities and randomness of the market trade time series. We don't study why market trade is random but describe how market randomness determines the stochasticity of price, return, and other economic variables. The description of random trade gives a firm basis for the description of averages and volatilities of market price and return.

We consider the randomness of market trade values and volumes as the origin of price and return stochasticity and describe how averages and volatilities of price and return depend on averages, volatilities, and correlations of market trade values and volumes. We call it the market-based approach to probability of price and return. In some sense, the uncertainty of macroeconomic variables is described by statistical moments of market trade. We argue that economic theory, to a large extent, is a description of trade statistical moments.

The rest of the paper is as follows: In Section 2, we discuss approximations of economic processes. In Section 3, we describe the composition of economic variables during some time interval. In Sections 4 and 5, we describe the market-based statistical moments of price and return. After these introductory sections, in Section 6, we consider the relations between economic theory and market trade statistical moments, introduce the secondary averaging procedure, and discuss links between macroeconomic variables and trade statistical moments. In Section 7, we discuss some practical outcomes. Section 8 – Conclusion. We are sure that readers are familiar with common use models of price and return probabilities, have skills in the use of statistical moments, etc., and know or can find on their own the definitions, notions, and terms that are not given in the text.

## 2. General considerations

To study economic theory, one should identify the main elements that compose the approximation of the economic relations. We consider economic agents as the elementary bricks that establish the economic system as a whole. As agents, we consider international companies, large banks, hedge funds, small firms, shops, households, etc. We assume that agents have many economic and financial variables like income and consumption, taxes and production, investment and profits, etc. All macroeconomic variables that define the evolution of the economic system as a whole are determined by the aggregations of corresponding variables of economic agents, or depend on them (Fox et al., 2017).

We consider market trade as the main factor that results in the change of agents' variables, macroeconomic variables, and the evolution of the economy as a whole. Agents make deals with other agents, and these trades change agents' variables and result in the change of macroeconomic variables. However, the frequency of trades in stock markets, FOREX, commodities markets, etc. is very high. A time interval *ε* between market trades can be equal to a second or even a fraction of a second. Such a high frequency of trades results in highly irregular trade time series and irregular changes of the corresponding agents' variables. In addition, high-frequency time series are of little help for the description of long-term macroeconomic relations. The collisions between the high frequency of market trades and the description of long-term relations raise the problem of the choice of the averaging time interval *Δ* that determines the time axis division of the macroeconomic model. Indeed, a high-frequency time series of market trades at times $t_i$ such as:

$$t_{i+1} - t_i = \varepsilon \quad ; \quad i = 1, 2.. \quad (2.1)$$

introduces the initial market time axis division multiple of *ε*. However, this initial market time axis division is too precise for macroeconomic modeling. To describe macroeconomic relations at a long horizon *T>> ε,* one should roughen the initial time axis division. To do that, one should choose the time interval *Δ,* such as *ε<< Δ <T*, and aggregate or average the market trade time series during *Δ*. The choice of the interval *Δ* and the aggregation of market trade time series during *Δ* smooths the irregularity of the initial trade time series and determines the collective impact of market trade on agents' and macroeconomic variables.

Any approximations of real economic processes or economic models require the choice of a time-averaging interval *Δ*. The duration of the interval *Δ* significantly determines the properties, reliability, and stability of the economic models and predictions that are based on variables aggregated during *Δ*. The sequence of averaging time intervals *Δ<Δ₂<Δ₃<..*

determines the sequence of economic models with more and more smooth changes of variables.

## 3. Properties of aggregated variables

Investment decisions during days or weeks require the averaging interval *Δ* no longer than days or weeks. The description of markets under decisions during hours requires the interval *Δ* no longer than hours. Long-term economic forecasts on the horizon of months or years could require the interval *Δ* to equal at least a month. Different durations of *Δ* result in different approximations of economic variables.

Let us consider how the averaging interval *Δ* determines the properties of collective economic variables. We start with a simple model and consider market trade with a particular asset. As such, one can consider stocks of a company, oil, metals, gold, FOREX trading, etc. We assume that the frequency of trades is determined by (2.1) and that *Δ* defines *Δ_k*:

$$\Delta_k = \left[t_k - \frac{\Delta}{2}; t_k + \frac{\Delta}{2}\right] \; ; \; t_k = t_0 + \Delta \cdot k \; ; \; k = 0, \pm 1, \pm 2, \ldots \quad (3.1)$$

We use *k=1,2,..* to describe averaging intervals in the future and *k=-1, -2, ..* in the past. We renumber the time series $t_i$ in such a way that $t_{ik}$ belongs to the intervals *Δ_k*:

$$t_{ik} \in \Delta_k \; ; \; i = 1, 2, \ldots N \quad (3.2)$$

The interval *Δ* substitutes the initial market time axis division $t_i$ (2.1) that is a multiple of *ε* with a new one, $t_k$ (3.1), that is a multiple of *Δ*. To express the change of economic variables as a result of trades during *Δ*, let us consider market trade value $C(t_{ik})$ and volume $U(t_{ik})$ at a time $t_{ik}$. During the interval *Δ*, the total trade value $C_Δ(t_k)$ and volume $U_Δ(t_k)$ take the form:

$$C_\Delta(t_k; 1) = \sum_{i=1}^{N} C(t_{ik}) \; ; \; U_\Delta(t_k; 1) = \sum_{i=1}^{N} U(t_{ik}) \quad (3.3)$$

Relations (3.3) define the collective change of market trade value *C_Δ(t_k;1)* and volume *U_Δ(t_k;1)* during *Δ*. Obviously, the time series of the trade value *C_Δ(t_k)* and volume *U_Δ(t_k)* at time *t_k, k=0,1,..* demonstrate more smooth dynamics than the initial high frequency and irregular market time series of the trade value *C(t_{ik})* and volume *U(t_{ik})* during *Δ_k* (3.1; 3.2). We use index 1 in (3.3) to highlight that the sums (3.3) are taken over the 1[st] degree of the variables on the right side. That index will play an important role in our further consideration, and we highlight its importance now. The duration of the interval *Δ* can be equal to a day, a week, a month, or a quarter, and that introduces the change of macroeconomic variables during the corresponding time interval. One can consider any market trades that are performed in the economy during the interval *Δ* alike to collective trade value and volume (3.3). The perfect methodology for aggregation of economic variables as sums of transactions that are made during the averaging interval *Δ* is presented by Fox et al. (2017). It describes the procedures that are in use for assessments of the official statistics of the National Accounts for collecting additive economic variables and subsequent assessments of non-additive variables such as prices, returns, inflation, etc. The use of collective variables, which are alike to (3.3), opens the way for modeling the dynamics of monthly, quarterly, or annual investments and sales, consumption and production, profits or expenses, etc. Most macroeconomic theories (Leontief, 1955; Sargent, 1979; Blaug, 1985; Romer, 1996; Krueger, 2002) describe relations between economic variables, which are composed of corresponding economic transactions during the selected interval $\Delta$. Obviously, macroeconomic theories are not limited to using only variables similar to (3.3). They use price, return, rates, indices, etc. Each market deal with the trade value $C(t_{ik})$ and volume $U(t_{ik})$ at time $t_{ik}$ defines the price $p(t_{ik})$ due to a trivial equation:

$$C(t_{ik}) = p(t_{ik})U(t_{ik}) \qquad (3.4)$$

For convenience, we consider all prices adjusted to the current value at time $t_0$. The predictions of price are the core problems of financial economics and generate an endless row of studies (Muth, 1961; Sharpe, 1964; Fama, 1965; Black and Scholes, 1973; Merton, 1973; Friedman, 1990; Cochrane and Hansen, 1992; Cochrane, 2001; Campbell, 2018). These references present only a negligible part of the asset pricing studies. At least since Bachelier (1900), the description of price as a random variable has become the most conventional: "in fact, the first author to put forward the idea to use a random walk to describe the evolution of prices was Bachelier" (Shiryaev, 1999). The forecasts of price and return probabilities at horizon $T$ are among the most studied problems of modern finance. However, the hidden economic barriers almost prohibit any exact predictions of price probabilities. The notion of price probability itself has multiple treatments. Below, we consider the market-based probabilities of price and return that depend on the randomness of the market trade.

## 4. The market-based statistical moments of price

The conventional treatment of price probability proposes that during $\Delta$ all $N$ trades have equal probabilities and the probability $P(p)$ of price $p$ is proportional to the frequency $m_p/N$ of trades at price $p$:

$$P(p) \sim m_p/N \qquad (4.1)$$

For convenience, we note (4.1) as the frequency-based approach to price probability that, during the last century, has been studied in a great number of papers. Researchers checked up almost all standard random distributions (Forbes et al., 2011; Walck, 2011) to verify their adequacy to different assessments of price probability. The frequency-based treatment (4.1) of price probability completely follows contemporary probability theory (Shiryaev, 1999; Shreve, 2004) and uses modern methods of probability theory in financial economics.

However, the conventional frequency-based approach (4.1) to price probability almost 35 years ago was supplemented by an assessment of the average price that takes into account the size of volumes of market deals and is well known as volume weighted average price (VWAP) (Berkowitz et al., 1988; Duffie and Dworczak, 2018). Using relations (3.1-3.4), VWAP $p(t_k;1)$ during the interval $\Delta_k$ takes the form:

$$p(t_k;1) = \frac{1}{\sum_{i=1}^{N} U(t_{ik})} \sum_{i=1}^{N} p(t_{ik}) U(t_{ik}) = \frac{C_\Delta(t_k;1)}{U_\Delta(t_k;1)} \qquad (4.2)$$

Relation (4.2) determines the average price $p(t_k;1)$ at time $t_k$ during the interval $\Delta_k$ (3.1; 3.2). The average price $p(t_k;1)$, or as it is called, the 1$^{st}$ statistical moment of price, is not enough to determine the price probability. To describe the probability of price as a random variable (Shiryaev, 1999; Shreve, 2004), one should determine all statistical moments $p(t_k;n)$ of price:

$$p(t_k;n) = E[p^n(t_{ik})] \quad ; \quad n = 1,2,\ldots \qquad (4.3)$$

We use $E[..]$ to note mathematical expectation during $\Delta_k$. The finite set of price statistical moments $p(t_k;m)$, $m=1,\ldots n$, defines an approximation of the price probability. The frequency-based probability (4.1) is a completely correct description of the random price (Shiryaev, 1999; Shreve, 2004) if one studies only irregular price time series during $\Delta_k$.

However, the market price $p(t_{ik})$ is a result of market trade (3.4). To consider the price as a random variable that is determined by (3.4), one should take into account the random properties of trade value $C(t_{ik})$ and volume $U(t_{ik})$ (Olkhov, 2021a; 2021b; 2022a; 2023a; 2023b). Indeed, equation (3.4) states that the random properties of trade value $C(t_{ik})$ and volume $U(t_{ik})$ determine the properties of the random price $p(t_{ik})$. We consider the price (3.4) as the result of market deals and describe the dependence of the market-based statistical moments of price on the statistical moments of market trade value $C(t_{ik})$ and volume $U(t_{ik})$. We call that the market-based approach to price probability. To support that simple proposal, we refer to Fox et al. (2017), who present the methodology for the assessment of aggregate price and other non-additive macroeconomic variables as a result of the aggregation of additive economic variables. The time series of the trade value $C(t_{ik})$ and volume $U(t_{ik})$ are examples of additive economic variables, and the sums of the trade value $C_\Delta(t_k)$ and volume $U_\Delta(t_k)$ (3.3) during the interval $\Delta_k$ determine the VWAP $p(t_k;1)$ (4.2).

In total, there is no single solution, no single rule, or law that determines a single definition of the price probability. The adherents of empirical evidence supporting any theoretical results and conclusions in economics and finance should be disappointed and discouraged. No econometric data exists that can provide any empirical evidence in favor of a frequency-based or market-based approach to the definition of price probability. Both of these two different considerations of price probability exist simultaneously. The frequency-based look is a generally accepted, conventional approach that models price probability under the assumption that all trade volumes $U(t_{ik})$ during $\Delta$ are constant. The market-based approach to price probability that we propose takes into account the randomness of trade values and volumes during $\Delta$ and establishes the direct dependence of price and return statistical moments on statistical moments and correlations on market trade values and volumes. That links market-based price probability with market randomness and macroeconomic evolution.

In this paper, we describe the market-based statistical moments of price and return and highlight their mutual relations with problems of economic and financial theory. We show that on the one hand, the market-based approach to price and return probabilities is determined by the stochasticity of market trade, and on the other hand, it highlights the origin of the successive approximations of the economic and financial theories.

We consider the irregular time series of trade value $C(t_{ik})$ and volume $U(t_{ik})$ during the interval $\Delta_k$ as the only source of the random properties of price and return. We consider the trade value and volume as random variables during $\Delta_k$ and assess their statistical moments by conventional frequency-based probability (4.1). We denote the *n*-th statistical moments of trade value $C(t_k;n)$ and volume $U(t_k;n)$ using frequency-based probability:

$$C(t_k;n) = E[C^n(t_{ik})] \sim \frac{1}{N}\sum_{i=1}^{N} C^n(t_{ik}) \quad ; \quad U(t_k;n) = E[U^n(t_{ik})] \sim \frac{1}{N}\sum_{i=1}^{N} U^n(t_{ik}) \quad (4.4)$$

We use the symbol "~" to highlight that (4.4) are the estimates of the n-th statistical moments during $\Delta_k$ by a finite number $N$ of terms of time series.

Let us take the total trade value $C_\Delta(t_k;1)$ and volume $U_\Delta(t_k;1)$ (3.3) during $\Delta_k$ (3.1; 3.2) and introduce the similar variables as sums of the *n*-th degree of trade value $C_\Delta(t_k;n)$ and volume $U_\Delta(t_k;n)$ during $\Delta_k$:

$$C_\Delta(t_k;n) = N \cdot C(t_k;n) = \sum_{i=0}^{N} C^n(t_{ik}) \; ; \; U_\Delta(t_k;n) = N \cdot U(t_k;n) = \sum_{i=0}^{N} U^n(t_{ik}) \quad (4.5)$$

Let us consider the n-th degree of the price equation (3.4):

$$C^n(t_{ik}) = p^n(t_{ik})U^n(t_{ik}) \quad ; \quad n = 1,2,\ldots \quad (4.6)$$

The equations (4.6) generate the set of the price statistical moments $p(t_k;m,n)$ in a form that is similar to the form of VWAP:

$$p(t_k;n,m) = \frac{1}{\sum_{i=1}^{N} U^m(t_{ik})} \sum_{i=1}^{N} p^n(t_{ik})U^m(t_{ik}) = \sum_{i=1}^{N} p^n(t_{ik})w(t_{ik};m) \quad (4.7)$$

The functions $w(t_{ik};m)$ in (4.7; 4.8) have the meaning of weighted functions but not price probabilities:

$$w(t_{ik}; m) = \frac{U^m(t_{ik})}{\sum_{i=1}^{N} U^m(t_{ik})} \quad ; \quad \sum_{i=1}^{N} w(t_{ik}; m) = 1 \; ; \; m = 1, 2, \ldots \qquad (4.8)$$

For each $m=1,2,\ldots$, the weighted functions define the sequence of price statistical moments $p(t_k;n,m)$ averaged during $\Delta_k$ (3.1; 3.2). In particular, the average of price squares, or the 2$^{nd}$ statistical moment $p(t_k;2,2)$:

$$p(t_k; 2,2) = \frac{1}{\sum_{i=1}^{N} U^2(t_{ik})} \sum_{i=1}^{N} p^2(t_{ik}) U^2(t_{ik}) = \sum_{i=1}^{N} p^2(t_{ik}) w(t_{ik}; 2) = \frac{C_\Delta(t_k;2)}{U_\Delta(t_k;2)} \qquad (4.9)$$

The 2$^{nd}$ price statistical moment $p(t_k;2,2)$ (4.9) is determined by the weighted function $w(t_{ik};2)$ (4.8) and matches (4.6). As follows from (4.9), $p(t_k;2,2)$ equals the ratio of the sum of squares of trade values $C_\Delta(t_k;2)$ to the sum of squares of trade volumes $U_\Delta(t_k;2)$ (4.5). We use the sequence of statistical moments (4.7) to define market-based price statistical moments. To do that, one should harmonize each next market-based price statistical moment with previous ones. As the market-based average price or the 1$^{st}$ statistical moment we take VWAP (4.2). To distinguish market-based price statistical moments from (4.7) we denote them as $a(t_k;n)$. We define $a(t_k;1)$ (4.10):

$$a(t_k; 1) = p(t_k; 1,1) = \frac{1}{\sum_{i=1}^{N} U(t_{ik})} \sum_{i=1}^{N} p(t_{ik}) U(t_{ik}) = \frac{C_\Delta(t_k;1)}{U_\Delta(t_k;1)} = \frac{C(t_k;1)}{U(t_k;1)} \qquad (4.10)$$

To define the 2$^{nd}$ market-based price statistical moment $a(t_k;2)$, the market-based price volatility $\sigma^2(t_k)$ (4.11) be non-negative. We use $E_m[\ldots]$ to denote the mathematical expectation of market-based price probability to distinguish it from the frequency-based mathematical expectation $E[\ldots]$. We define market-based price volatility $\sigma^2(t_k)$ as:

$$\sigma^2(t_k) = E_m[(p(t_{ik}) - a(t_k; 1))^2] = a(t_k; 2) - a^2(t_k; 1) \geq 0 \qquad (4.11)$$

To determine volatility $\sigma^2(t_k)$ (4.11) we set it to be equal to:

$$\sigma^2(t_k) = a(t_k; 2) - a^2(t_k; 1) = \sum_{i=1}^{N} [p(t_{ik}) - a(t_k; 1)]^2 w(t_{ik}; 2) \qquad (4.12)$$

The relations (4.12) define the market based price volatility $\sigma^2(t_k)$ (4.13) and the 2$^{nd}$ price statistical moment $a(t_k;2)$ (4.14). We refer to Olkhov (2021a; 2021b; 2022) for further details:

$$\sigma^2(t_k) = \frac{\Omega_C^2(t_k) + a^2(t_k;1)\Omega_U^2(t_k) - 2a(t_k;1)corr\{C(t_k)U(t_k)\}}{U(t_k;2)} \qquad (4.13)$$

$$a(t_k; 2) = \frac{C(t_k;2) + 2a^2(t_k;1)\Omega_U^2(t_k) - 2a(t_k;1)corr\{C(t_k)U(t_k)\}}{U(t_k;2)} \qquad (4.14)$$

The market-based price volatility $\sigma^2(t_k)$ (4.13) and the 2$^{nd}$ price statistical moment $a(t_k;2)$ (4.14) depend on trade value volatility $\Omega_C^2(t_k)$ and volume volatility $\Omega_U^2(t_k)$ (4.15)

$$\Omega_C^2(t_k) = C(t_k; 2) - C^2(t_k; 1) \; ; \quad \Omega_U^2(t_k) = U(t_k; 2) - U^2(t_k; 1) \qquad (4.15)$$

and on correlation $corr\{C(t_k)U(t_k)\}$ (4.16) between trade value and volume:

$$corr\{C(t_k)U(t_k)\} = E\big[(C(t_{ik}) - C(t;1))(U(t_{ik}) - U(t;1))\big] = E[C(t_{ik})U(t_{ik})] - C(t_k;1)U(t_k;1) \quad (4.16)$$

The joint average $E[C(t_k)U(t_k)]$ (4.17) of the product of trade value and volume takes the form:

$$E[C(t_k)U(t_k)] = \frac{1}{N}\sum_{i=1}^{N} C(t_{ik})U(t_{ik}) \quad (4.17)$$

We reduce derivation of the market-based statistical moments by the first two and will discuss this limit below.

## 5. The market-based average and volatility of return

We use the market-based approach to derive average and volatility of return (Olkhov, 2023a; 2023b). Let us choose a time shift $\tau$ and consider return $r(t_{ik},\tau)$ as a ratio of price $p(t_{ik})$ at time $t_{ik}$ to price $p(t_{ik}-\tau)$ at a time $t_{ik}-\tau$ in the past:

$$r(t_{ik}, \tau) = \frac{p(t_{ik})}{p(t_{ik}-\tau)} \quad (5.1)$$

We take the time shift $\tau$ to be a multiple of $\varepsilon$, and hence, the time series $t_{ik}-\tau$ belongs to some past intervals $\Delta_j$, $j=k-m$. Let us transform equation (3.4) as follows:

$$C(t_{ik}) = \frac{p(t_{ik})}{p(t_{ik}-\tau)} p(t_{ik} - \tau)U(t_{ik}) = r(t_{ik},\tau)C_o(t_{ik},\tau) \quad (5.2)$$

$$C_o(t_{ik}, \tau) = p(t_{ik} - \tau)U(t_{ik}) \quad (5.3)$$

The relations (5.3) define the value $C_o(t_{ik},\tau)$ of the trade volume $U(t_{ik})$ at a price $p(t_{ik}-\tau)$ in the past. The *n-th* degree of (5.2) defines equation (5.4) on the n-th degree of return $r^n(t_{ik},\tau)$, similar to equation on price $p(t_{ik})$ (4.6).

$$C^n(t_{ik}) = r^n(t_{ik},\tau)C_o^n(t_{ik},\tau) \quad (5.4)$$

Actually, the market-based average return $h(t_k;1)$ was introduced by Markowitz (1952) in a form that is almost similar to the definition of VWAP. Markowitz (1952) defines portfolio return as return "weighted with weights equal the relative amount invested in security." One can consider market deals at times $t_{ik}$ during the interval $\Delta_k$ as a portfolio and estimate the market-based average "anticipated" return with a time shift $\tau$ due to Markowitz's definition. It almost completely reproduces the definition of VWAP but substitutes the trade volumes with the past values of the trades during $\Delta_k$. At the same time, one can consider VWAP as the usual assessment of the portfolio's average price as the ratio of the total value of stocks to the total volume of stocks in the portfolio. Using (4.5; 5.3; 5.4), one can define Markowitz's average return that we consider as the market-based average return $h(t_k,\tau;n)$ during $\Delta_k$ similar to the market-based average price $a(t_k;1)$ (4.10):

$$h(t_k,\tau;1) = \frac{1}{\sum_{i=1}^{N} C_o(t_{ik},\tau)} \sum_{i=1}^{N} r(t_{ik},\tau)C_o(t_{ik},\tau) = \frac{C_\Delta(t_k;1)}{C_{o\Delta}(t_k,\tau;1)} = \frac{C(t_k;1)}{C_o(t_k,\tau;1)} \quad (5.5)$$

$$C_{o\Delta}(t_k,\tau;n) = NC_o(t_k,\tau;n) = \sum_{i=1}^{N} C_o^n(t_{ik},\tau) \quad (5.6)$$

The definitions of market-based volatility of return $v^2(t_k,\tau)$ and the 2nd statistical moment $h(t_k,\tau;2)$ are almost similar to (4.13;4.14). We refer to Olkhov (2023a-2023c) for details.

$$v^2(t_k,\tau) = \frac{\Omega_C^2(t_k)+h^2(t_k,\tau;1)\Phi^2(t_k,\tau)-2h(t_k,\tau;1)corr\{C(t_k)C_o(t_k,\tau)\}}{C_o(t_k,\tau;2)} \quad (5.7)$$

$$h(t_k,\tau;2) = \frac{C(t_k;2)+2h^2(t_k,\tau;1)\Phi^2(t_k,\tau)-2h(t_k,\tau;1)corr\{C(t_k)C_o(t_k,\tau)\}}{C_o(t_k,\tau;2)} \quad (5.8)$$

In (5.7; 5.8) we denote volatility $\Phi(t,\tau)$ (5.9) of the past trade value:

$$\Phi^2(t_k,\tau) = E[(C_o(t_{ik},\tau)-C_o(t_k,\tau;1))^2] = C_o(t_k,\tau;2)-C_o^2(t_k,\tau;1) \quad (5.9)$$

The correlation $corr\{C(t_k)C_o(t_k,\tau)\}$ between the current and past trade values takes the form:

$$corr\{C(t_k)C_o(t_k,\tau)\} = E[C(t_{ik})C_o(t_{ik},\tau)] - C(t_k;1)C_o(t_k,\tau;1) \quad (5.10)$$

The joint average $E[C(t_{ik})C_o(t_{ik},\tau)]$ of the product of the current and past trade values:

$$E[C(t_{ik})C_o(t_{ik},\tau)] = \frac{1}{N}\sum_{i=1}^{N} C(t_{ik})C_o(t_{ik},\tau) \quad (5.11)$$

The relations (5.5; 5.7; 5.8) define the market-based average $h(t_k,\tau;)$, volatility $v^2(t_k,\tau)$, and the 2nd statistical moment $h(t_k,\tau;2)$ of return. These relations highlight the dependence on statistical moments (4.5; 5.6), volatilities (4.15; 5.9) and correlations (4.16; 5.10) of the market trade value, volume, and the past value (Olkhov, 2021a-2023c).

## 6. Economic theory and statistical moments

Now we discuss the main matter of this paper: the links between economic theory and the statistical moments of economic variables. As usual, economic models describe the mutual dependence of macroeconomic variables that results in the change of variables. The change of each macroeconomic variable is determined by the changes of corresponding variable of agents during a particular time interval $\Delta$. For example, the change of macroeconomic investments, credits, taxes, etc. is determined by the change of the corresponding variables of economic agents during $\Delta$. In turn, agents' market deals during $\Delta$ cause a change of agents' variables. In total, the change of macroeconomic variables during $\Delta$ is determined by market deals during $\Delta$. We highlight that economic and financial transactions of agents are the only processes that change agents' variables. Any change of macroeconomic investments, credits, GDP, etc. during the interval $\Delta$ occurs only as a result of agents' transactions. Almost all macroeconomic variables are composed of sums of the 1st degrees of market trade values and volumes. For example, the change of macroeconomic investment is determined by the sum of the investment deals of agents during $\Delta$. For convenience, we call them the 1st degree economic variables. We call economic models, which describe relations between the 1st degree variables as the 1st order economic theories or the 1st degree approximations.

However, economic and financial trades significantly depend on agents' expectations of future prices, returns, volatilities, etc. Predictions of market-based volatilities of price and return depend on the forecasts of corresponding volatilities and correlations of market trade values and volumes (4.13; 5.7). Agents' expectations and predictions of price and return volatilities that impact agents' trade decisions depend on the forecasts of the 2$^{nd}$ statistical moments, volatilities, and correlations of the trade values and volumes (4.4; 4.15; 4.16; 5.6; 5.9; 5.10). In simple words, to describe macroeconomic variables that are composed of sums of market trade values and volumes, one should also model the relations between variables of the 2$^{nd}$ degree. The attempts to approximate the market-based price or return probabilities by the first *n* statistical moments require the development of economic theories that describe the relations between variables composed of sums of market trade values and volumes in the *m-th* degrees for *m=1,…n*. We believe that theoretical economics should be considered as successive approximations of the *n*-th statistical moments of trade, price, return, and economic variables starting with the modeling of the 1$^{st}$ degree economic variables, which are described by conventional economic models. Each next *n-th* approximation adds an additional layer of description that is formed of economic variables composed of the sums of the *n-th* degrees of market deals during the selected time averaging interval *Δ* (Olkhov, 2021b; 2023c). For each *n=1,2,...*, we consider the n-th approximation as the description of macroeconomic variables composed by the *m-th* degrees of market trades and the *m-th* statistical moments of trade, price and return for *m=1, ..n*.

The choice of the averaging time interval adds extra complexity to economic approximations. Indeed, one can consider the sequence of the averaging intervals *Δ< Δ$_2$< Δ$_3$<….* We call the transition from the economic approximation, which is determined by the interval *Δ*, to the approximation, which is determined by the interval *Δ$_2$ >> Δ*, the secondary averaging procedure. One can repeatedly average the sequence of economic models to obtain more and more averaged approximations. Actually, the averages can demonstrate irregular or random behavior during a long interval *Δ$_2$>>Δ*. The average values *C(t$_k$;1)*, volumes *U(t$_k$;1)*, market-based average price *a(t$_k$;1)*, and return *h(t$_k$,τ;1)* could behave irregularly or randomly during the interval *Δ$_2$ >> Δ*. Assume that:

$$\Delta_2 = M \cdot \Delta \quad ; \quad M \gg 1 \tag{6.1}$$

$$t_k \in \left[t - \frac{\Delta_2}{2}; t + \frac{\Delta_2}{2}\right] \quad ; \quad k = 0, 1, 2, \ldots M \tag{6.2}$$

Due to (4.10) and similar to (3.4) one can define the average trade price equation (6.3):

$$C(t_k; 1) = a(t_k;)U(t_k; 1) \quad ; \quad C^n(t_k; 1) = a^n(t_k;)U^n(t_k; 1) \tag{6.3}$$

Respectively, the market-based average return (5.5) defines the equation (6.4):

$$C(t_k;1) = h(t_k,\tau;1)U(t_k;1) \quad ; \quad C^n(t_k;1) = h^n(t_k,\tau;1)U^n(t_k;1) \tag{6.4}$$

Let us consider the average trade values $C(t_k;1)$, volumes $U(t_k;1)$ (4.5; 6.4), and past values $C_o(t_k,\tau;n)$ (6.4), as random variables for $t_k$, $k=1,2,..M$ during the interval $\Delta_2$ (6.1) and define:

$$C_\Delta(t;1,n) = M \cdot C(t;1,n) = \sum_{k=0}^{M} C^n(t_k;1) \tag{6.5}$$

$$U_\Delta(t;n) = M \cdot U(t;1,n) = \sum_{k=0}^{N} U^n(t_k,1) \tag{6.6}$$

$$C_{o\Delta}(t,\tau;1,n) = M \cdot C_o(t,\tau;1,n) = \sum_{k=1}^{M} C_o^n(t_{ik},\tau;1) \tag{6.7}$$

Then the secondary market-based average price $a_2(t;1)$, takes the form:

$$a_2(t;1) = \frac{1}{\sum_{k=1}^{M} U(t_k;1)} \sum_{k=1}^{M} a(t_k;1)U(t_k;1) = \frac{C_\Delta(t;1,1)}{U_\Delta(t;1;1)} = \frac{C(t;1,1)}{U(t;1.1)} \tag{6.8}$$

The price $a_2(t;1)$ has a form of VWAP and describes the market-based averaging of irregular average prices $a(t_k;1)$ during a long averaging interval $\Delta_2 \gg \Delta$ taking into account the random size of average volumes $U(t_k;1)$ (4.5), or what is equal, the irregular sums of volumes $U_\Delta(t_k;1)$ (4.5). We denote the secondary market-based volatility $\sigma^2(t;1)$ (6.9) to highlight that (6.9) describes the volatility of random fluctuations of the market-based average price $a(t_k;1)$ during the interval $\Delta_2 \gg \Delta$:

$$\sigma^2(t_k;1) = \frac{\Omega_{C2}^2(t) + a_2^2(t;1)\Omega_{U2}^2(t) - 2a_2(t;1)corr\{C(t;1,1)U(t;1,1)\}}{U(t;1,2)} \tag{6.9}$$

We underline the difference between volatility $\sigma^2(t;1)$ (6.9) of fluctuating average price $a(t_k;1)$ and averaging of the irregular time-series of price volatility $\sigma^2(t_k)$ (4.13). Those are different variables. The secondary volatilities $\Omega_{C2}^2(t)$ and $\Omega_{U2}^2(t)$ (6.10) of average trade values $C(t_k;1)$ and volumes $U(t_k;1)$ during $\Delta_2 \gg \Delta$ take the form similar to (4.15):

$$\Omega_{C2}^2(t) = C(t;1,2) - C^2(t;1,1) \quad ; \quad \Omega_U^2(t) = U(t;1,2) - U^2(t;1,1) \tag{6.10}$$

The correlation $corr\{C(t;1,1)U(t;1,1)\}$ (6.11) between the average trade values $C(t_k;1)$ and volumes $U(t_k;1)$ takes the form similar to (4.16):

$$corr\{C(t;1,1)U(t;1,1)\} = E[C(t_k;1)U(t_k;1)] - C(t;1,1)U(t;1,1) \tag{6.11}$$

$$E[C(t_k;1)U(t_k;1)] = \frac{1}{M}\sum_{k=1}^{M} C(t_k;1)U(t_k;1) \tag{6.12}$$

The secondary market-based average return $h_2(t,\tau;1)$ (6.13) describes the market-based averaging of the irregular time series of the average return $h(t_k,\tau;1)$ during $\Delta_2 \gg \Delta$:

$$h_2(t,\tau;1) = \frac{1}{\sum_{k=1}^{M} C_o(t_k,\tau;1)} \sum_{k=1}^{M} h(t_k,\tau;1)C_o(t_k,\tau;1) = \frac{C_\Delta(t;1)}{C_{o\Delta}(t,\tau;1)} = \frac{C(t;1)}{C_o(t,\tau;1)} \tag{6.13}$$

The secondary market-based volatility $v^2(t_k,\tau;1)$ (6.14) of random fluctuations of the average return $h(t_k,\tau;1)$ during $\Delta_2 \gg \Delta$ take the form alike to (5.7):

$$v^2(t,\tau;1) = \frac{\Omega_{C2}^2(t) + h_2^2(t,\tau;1)\Phi_2^2(t,\tau) - 2h_2(t,\tau;1)corr\{\{C(t;1,1)C_o(t;1,1)\}}{C_o(t;1,2)} \tag{6.14}$$

$$\Phi_2^2(t,\tau) = C_o(t,\tau;1,2) - C_o^2(t,\tau;1,1) \qquad (6.15)$$

$$corr\{C(t;1,1)C_o(t,\tau;1,1)\} = E[C(t_k;1)C_o(t_k,\tau;1)] - C(t;1,1)C_o(t,\tau;1,1) \qquad (6.16)$$

$$E[C(t_k;1)C_o(t_k,\tau;1)] = \frac{1}{M}\sum_{k=1}^{M} C(t_k;1)C_o(t_k,\tau;1) \qquad (6.17)$$

The use of successive time averaging intervals $\Delta<\Delta_2<\Delta_3<....$ introduces repeated averaging procedures of the statistical moments and correlations of the trade values (6.5; 6.7), volumes (6.6), price (6.8-6.12), and return (6.13-6.17). Each next interval $\Delta_j$, $j=2,3,..$ introduces an averaging procedure over the trade values and volumes, which were averaged over the previous interval $\Delta_{j-1}$. Moreover, each interval $\Delta_j$ introduces the next approximation of macroeconomic variables that are determined by the market trade values and volumes averaged over $\Delta_j$. The sequence of time averaging intervals $\Delta<\Delta_2<\Delta_3<....$ introduces the sequence of macroeconomic approximations. The second statistical moments of trade values $C(t;1,2)$ (6.5), $C_o(t,\tau;1,2)$ (6.7), and volumes $U(t;1,2)$ (6.6) introduce macroeconomic variables of the 2$^{nd}$ degree, which together with price (6.9) and return (6.14) volatilities establish the basis for the 2$^{nd}$ degree macroeconomic theories. The sequence of the averaging intervals, sums of different degrees of market trades, repeated averaging procedures, etc. generate a great hierarchy of macroeconomic approximations strongly linked to approximations of statistical moments of market trade, price, and return.

Economics is a complex system with strong forward-and-backward links and constraints. Obviously, the randomness during a selected time interval cannot be an exclusive property of market trade, price, and return. All agents' variables can demonstrate irregular or random behavior during a selected averaging interval, and their randomness in varying degrees is governed by the randomness of market trade. The duration of the averaging interval is the key factor for the description of the stochasticity, averages, and volatilities of macroeconomic variables. The successive approximations of macroeconomics generated by the sequence of the averaging intervals of different durations and by the hierarchy of the n-th statistical moments of market trades establish a rather complex hierarchy of economic models.

However, the models that describe the economic evolution of the variables of the 1$^{st}$ and 2$^{nd}$ degrees are absent now. Moreover, there is no econometric data and methodology that could help assess the current values of most 2$^{nd}$ degree macroeconomic variables. We believe that Douglas Fox and coauthors (Fox et al., 2017) can develop the "NIPA Handbook-2", which would describe the 2$^{nd}$ degree assessments of the National Accounts determined by the sums of squares of corresponding economic transactions during an averaging interval $\Delta$. That would open the doors for the development of macroeconomic models of the 2$^{nd}$ degree.

Economic and financial theories describe macroeconomic variables and the statistical moments of market trade, price, and return. Economic randomness is a property of a particular averaging interval. The sequence of averaging intervals generates the sequence of economic theories as a set of successive approximations of statistical moments. Each next level of the approximation models an extra degree of statistical moments of the market trade, price, return, and economic variables. The recognition of the fact that the description of macroeconomic variables depends on the market-based statistical moments of trade, price, and return highlights the impact of probability on theoretical economics.

## 7. Practical outcomes

Our look at theoretical economics as a set of successive approximations, each of which describes the extra n-th degree of economic variables and statistical moments, could seem too abstract, but it has important practical applications. We briefly consider some.

In the coming years, in the best case, the predictions of the market-based probabilities of price and return are limited by the first two statistical moments. The predictions of the volatility of price or return at a horizon $T$ require forecasting the volatilities and correlations of market trade values and volumes at the same horizon and the development of a $2^{nd}$ degree economic model. The predictions of the market-based volatilities of price and return, which neglect the $2^{nd}$ degree economic model, would have low economic justification and could be highly uncertain (Olkhov, 2021b; 2023c). The development of $2^{nd}$ degree economic models may take years, but it is the only way to improve the predictability and reliability of economic theory. Predictions of the higher statistical moments require the development of economic models of the $3^{rd}$ degree, $4^{th}$ degree, etc. Until then, the forecasts of price and return probabilities are limited at best by the predictions of the first two statistical moments. Thus, Gaussian-type probabilities of price and return are the only possible predictions for many years to come.

The second problem concerns the reliability of the most common risk-hedging tool, Value-at-Risk (VaR). Indeed, the use of VaR (Longerstaey and Spencer, 1996; Duffie and Pan, 1997; Tobias and Brunnermeier, 2016) is based on the predictions at a horizon $T$ of the price or return probabilities determined by the conventional frequency-based approach (4.1). The heart of VaR is the assessment of the integrals of the left tails of the forecasts of probabilities as benchmarks for possible losses. Basically, the users of VaR are large banks, investment funds, financial institutions, etc. that manage hundreds of billions of USD of assets, stocks, and funds. To protect the value of their assets from a decline in price or return, they estimate possible losses through assessments of the left-tail integrals of the predicted probabilities. The main trouble concerns the choice of probabilities. The companies that manage hundreds of billions of USD must care about the randomness of market trades and hence should assess the market-based price and return probabilities that are determined by the statistical moments and correlations of trade values and volumes. Frequency-based assessments of price or return probabilities ignore the dependence on the random size of trade volumes. The differences between the market-based and the frequency-based approaches to price and return probabilities are the source of the additional risks and extra losses.

However, even the use of market-based probabilities of price and return carries hidden complexities. As we discussed above, the current state of economic theory at best limits the accuracy of any predictions of probabilities of price and return by Gaussian distributions. These are not very precise predictions. In total, the concept of VaR confirms the elementary thesis: no methods exist that can overcome internal economic obstacles using surrogates, like VaR, that don't solve but simply neglect the essence of the economic barriers.

## 8. Conclusion

This paper illuminates the links between the description of macroeconomic variables and the statistical moments of market trade, price, and return. The change in macroeconomic variables during the averaging interval $\Delta$ is determined by the sums of the n-th degrees of economic trade values and volumes during $\Delta$. The description of the averages and volatilities of price and return depends on the averages, volatilities, and correlations of the market trade values and volumes. One could consider theoretical economics as a set of successive approximations of statistical moments of market trade, price, return, and economic variables. The tight links between the descriptions of statistical moments of trade, price, return, and other macroeconomic variables highlight the unexpected features of economic theory. This paper doesn't consider agents' expectations, economic legislation, economic and market risks and their influence on agents' trade decisions, and many other factors that, for sure, impact the evolution and randomness of market trade and economics as a whole. That would greatly complicate the economic theory, and we leave it for further studies.

# References

Bachelier, L., (1900). Théorie de la speculation, Ann. Scientif. l'É.N.S. 3e série, **17**, 21-86

Black, F. and M. Scholes, (1973). The Pricing of Options and Corporate Liabilities, J. Political Economy, 81 (3), 637-654

Berkowitz, S.A., Dennis, E., Logue, D.E., Noser, E.A. Jr. (1988). The Total Cost of Transactions on the NYSE, *The Journal of Finance*, 43, (1), 97-112

Blaug, M. (1985). Economic theory in retrospect, Cambridge Univ.Press, 4-th ed., 760

Campbell, J.Y., (2018). Financial Decisions and Markets: A Course in Asset Pricing, Princeton Univ. Press, NJ, 477

Cantillon, R. (1730). An Essay on Economic Theory, 1-243, Translated Saucier C., Ed. Thornton, M., (2010). The Ludwig von Mises Institute, 254

Cochrane, J.H. and L.P. Hansen, (1992). Asset Pricing Explorations for Macroeconomics. Ed., Blanchard, O.J., Fischer, S. NBER Macroeconomics Annual 1992, v. 7, 115 – 182

Cochrane, J.H. (2001). Asset Pricing. Princeton Univ. Press, Princeton, US

Duffie, D. and J. Pan, (1997). An Overview of Value-at-Risk, J. of Derivatives, 4, (3), 7-49

Duffie, D. and P. Dworczak, (2018). Robust Benchmark Design, NBER WP 20540, 1-56

Fama, E.F. (1965). The Behavior of Stock-Market Prices. J. Business, 38 (1), 34-105

Forbes, C., Evans, M., Hastings, N., and B. Peacock, (2011). Statistical Distributions. Wiley

Fox, D.R. et al. (2017). Concepts and Methods of the U.S. National Income and Product Accounts. BEA, US.Dep. Commerce, 1-447

Friedman, D.D. (1990). Price Theory: An Intermediate Text. South-Western Pub. Co., US

Greenwald, B. and J. E. Stiglitz, (1987). Keynesian, New Keynesian and New Classical Economics, Oxford Economic Papers, 39 (1) 119-133

Hicks, J.R., (1937). Mr. Keynes and the "Classics"; A Suggested Interpretation, Econometrica, 5 (2), 147-159

Krueger, D. (2002). Macroeconomic Theory, Stanford Univ., 294

Kurz, H.D. and N. Salvadori, (2003). Classical economics and modern theory: studies in long-period analysis, Routledge, 325

Leontief, W., (1955). Some Basic Problems of Empirical Input-Output Analysis, 9-52, in Ed. Goldsmit,R.W., Input-Output Analysis: An Appraisal, Princeton Univ.Press, 369

Leontief, W., (1973). Structure of the World Economy, Nobel Memorial Lecture, 1-16

Longerstaey, J. and M. Spencer, (1996). RiskMetrics, Technical Document, 4-th Ed., Morgan Guaranty Trust Company NY, 296 p

Markowitz, H. (1952). Portfolio Selection, J. Finance, 7(1), 77-91

Merton, R.C. (1973). An Intertemporal Capital Asset Pricing Model, Econometrica, 41(5), 867-887

Muth, J.F. (1961). Rational Expectations and the Theory of Price Movements, Econometrica, 29, (3) 315-335

Neumann, J.V., (1945). A Model of General Economic Equilibrium, Rev. Econ. Studies, 13(1), 1-9

Olkhov, V. (2021a). Three Remarks On Asset Pricing, SSRN WP 3852261, 1-24

Olkhov, V., (2021b). Theoretical Economics and the Second-Order Economic Theory. What is it?, MPRA WP 110893, 1-14

Olkhov, V. (2022). The Market-Based Asset Price Probability, MPRA WP115382, 1-21

Olkhov, V., (2023a). The Market-Based Probability of Stock Returns, SSRN WP 4350975

Olkhov, V., (2023b). The Market-Based Statistics of 'Actual' Returns of investors, MPRA WP 116896, 1-16

Olkhov, V., (2023c). Economic Complexity Limits Accuracy of Price Probability Predictions by Gaussian Distributions, SSRN WP 4550635, 1-33

Romer, D. (1996). Advanced macroeconomics, McGraw-Hill, 550

Sargent, T.J. (1979). Macroeconomic Theory, Academic Press, 404

Sharpe, W.F. (1964). Capital Asset Prices: A Theory of Market Equilibrium under Conditions of Risk. The Journal of Finance, 19 (3), 425-442

Schumpeter, J.A. (1939). Business Cycles, McGraw-Hill, NY, 461

Shiryaev, A.N. (1999). Essentials Of Stochastic Finance: Facts, Models, Theory. World Sc. Pub., Singapore. 1-852

Shreve, S. E. (2004). Stochastic calculus for finance, Springer finance series, NY, USA

Solow, R.M., (1956). A Contribution to the Theory of Economic Growth, Quart. J. Economics, 70 (1), 65-94

Tobias, A. and M. K. Brunnermeier, (2016). CoVaR, Amer. Econ. Rev., 106(7), 1705–1741

Vines, D. and S. Wills (Ed.) (2018). Rebuilding Macroeconomic Theory, Oxford Rev. Econ. Policy, 34 (1–2), 353

Walck, C. (2011). Hand-book on statistical distributions. Univ.Stockholm, SUF–PFY/96–01

Wickens, M. (2008). Macroeconomic Theory, A Dynamic General Equilibrium Approach, Princeton Univ. Press, 489